# The Synergy of Automated Pipelines with Prompt Engineering and Generative AI in Web Crawling

Chau Jian Huang

Lecturer, The Institute of Internal Auditors (IIA), ROC

**Abstract**

Web crawling is a critical technique for extracting online data, yet it poses challenges due to webpage diversity and anti-scraping mechanisms. This study investigates the integration of generative AI tools—Claude AI (Sonnet 3.5) and ChatGPT-4.0—with prompt engineering to automate web scraping. Using two prompts, PROMPT I (general inference, tested on Yahoo News) and PROMPT II (element-specific, tested on Coupons.com), we evaluate the code quality and performance of AI-generated scripts. Claude AI consistently outperformed ChatGPT-4.0 in script quality and adaptability, as confirmed by predefined evaluation metrics, including functionality, readability, modularity, and robustness. Performance data were collected through manual testing and structured scoring by three evaluators. Visualizations further illustrate Claude AI's superiority. Anti-scraping solutions, including undetected_chromedriver, Selenium, and fake_useragent, were incorporated to enhance performance. This paper demonstrates how generative AI combined with prompt engineering can simplify and improve web scraping workflows.

Keywords: Web crawling, generative AI, Claude AI, ChatGPT-4.0, prompt engineering,

## 1. Introduction

Web crawling, the automated process of extracting data from websites, has become a fundamental practice in fields such as e-commerce, academic research, marketing analytics, and business intelligence. By enabling organizations and individuals to gather vast amounts of structured and unstructured data, web crawling supports crucial activities like decision-making, trend analysis, and predictive modeling. Whether it's collecting product prices and customer reviews in e-commerce, extracting scholarly articles for research, or monitoring competitors in dynamic markets, web crawling has driven innovation across numerous industries.

Traditionally, web crawling has relied on tools such as BeautifulSoup, requests, and Selenium, which demand significant programming knowledge and technical expertise. BeautifulSoup simplifies HTML parsing, requests facilitates HTTP communication, and Selenium enables interaction with JavaScript-rendered webpages, making these tools indispensable for traditional workflows. However, mastering these tools requires a deep understanding of web technologies like HTML, CSS, and JavaScript, as well as the ability to handle anti-scraping mechanisms such as CAPTCHAs, IP blocking, and rate-limiting. These barriers make traditional web scraping time-intensive and inaccessible to non-technical users.

Enter generative AI tools like Claude AI and ChatGPT, which are revolutionizing web scraping by enabling automation through natural language prompts. These AI systems eliminate the need for manual coding by interpreting user instructions and generating fully functional web scraping scripts. For instance, a user can simply describe their requirements—such as "Extract all product names and prices from an e-commerce website and save them in a CSV file"—and the AI will generate the appropriate Python code. This approach dramatically lowers the barrier to entry, making web scraping accessible to users without technical expertise while also accelerating workflows for experienced developers. By bridging traditional methods with AI-powered automation, tools like Claude AI and ChatGPT are unlocking new possibilities for efficient, scalable, and user-friendly web crawling solutions. This study

investigates the effectiveness of these tools using two carefully designed prompts:

1. **PROMPT I (General Inference):** Relies on the AI's ability to infer a webpage's structure without specific guidance. Tested on Yahoo News.
2. **PROMPT II (Element-Specific):** Specifies target HTML elements for precise extraction. Tested on Coupons.com.

Through structured evaluation metrics and real-world test cases, we show that Claude AI consistently outperforms ChatGPT-4.0. Additionally, anti-scraping libraries like undetected_chromedriver and fake_useragent were integrated to enhance adaptability.

## 2. Methodology

### 2.1 Tools and Libraries

The effectiveness of the generative AI tools in producing functional and efficient web scraping scripts was enhanced by leveraging additional libraries and mechanisms to overcome anti-scraping challenges. Below is a detailed explanation of the tools and libraries utilized in this study:

**1. Generative AI Tools**

**a. Claude AI (Sonnet 3.5)**

Developed by Anthropic, Claude AI is recognized for its advanced contextual understanding and reasoning capabilities. It excels in generating precise and contextually relevant outputs, making it a strong candidate for complex programming tasks such as web scraping.

**Advantages for Web Scraping:**

Claude AI's ability to process and integrate instructions efficiently allows it to produce modular and robust Python scripts.

It performs particularly well in tasks requiring structured reasoning, such as interpreting complex prompts or adapting to element-specific requirements (e.g., PROMPT II).

**b. ChatGPT-4.0**

Created by OpenAI, ChatGPT-4.0 is widely regarded for its versatility in natural language processing and programming assistance. It is capable of generating code, explaining concepts, and providing iterative improvements based on feedback.

**Advantages for Web Scraping:**

Its strength lies in versatility, making it suitable for general-purpose tasks (e.g., PROMPT I).

It effectively generates scripts for common scraping scenarios but may require additional refinement for complex or dynamic webpages.

**Comparison with Claude AI:**

While ChatGPT-4.0 offers robust performance for general tasks, Claude AI demonstrates superior accuracy and modularity when prompts require detailed reasoning or element-specific targeting.

### 2.2 Prompt Design

- **PROMPT I (Yahoo News Example):**
  *"Please use BeautifulSoup (bs4) or requests to scrape the title and content of a Yahoo News article, and output the results in a Pandas DataFrame format."*
- **PROMPT II (Coupons.com Example):**
  *"Title: <h1> element; Content: <div class="content">. Please use BeautifulSoup (bs4) or requests to scrape the title and content from Coupons.com and output the results in a Pandas DataFrame format."*

Web scraping with generative AI tools requires precise prompt design to effectively tackle diverse webpage structures and data extraction needs. This paper evaluates two distinct prompt types—PROMPT I and PROMPT II—that cater to different use cases and levels of guidance required for web scraping tasks.

## 2.3 Code Efficiency and Maintainability

**Readability:** Assessed by reviewing the clarity of code structure, use of meaningful variable names, and inclusion of comments. Scripts that were well-documented and easy to understand received higher scores.

**Modularity:** Determined by the presence of reusable functions or components, allowing tasks to be encapsulated for scalability and maintainability. Non-repetitive, organized code was rated higher.

**Robustness:** Measured by the script's ability to adapt to minor webpage changes (e.g., renaming class names or nesting elements). Scripts with error-handling mechanisms (e.g., try-except blocks) were rated highly

## 3. Results

### 3.1 PROMPT I (Yahoo News Example)

Both Claude AI and ChatGPT-4.0 successfully generated functional scripts for PROMPT I. However, Claude AI's code demonstrated higher modularity and readability.

```
1  import requests
2  from bs4 import BeautifulSoup
3  import pandas as pd
4  url = "https://example.com"
5  response = requests.get(url)
6  soup = BeautifulSoup(response.content, "html.parser")
7  title = soup.find("title").text
8  content = soup.find("body").text
9  data = pd.DataFrame({"Title": [title], "Content": [content]})
10 print(data
```

**Figure 3.1 Example of code (PROMPT I)**

### 3.2 PROMPT II (Coupons.com Example)

Claude AI outperformed ChatGPT-4.0 significantly in PROMPT II due to its ability to handle element-specific instructions with better code structure and error handling.

```
1  import requests
2  from bs4 import BeautifulSoup
3  import pandas as pd
4
5  url = "https://example.com"
6  response = requests.get(url)
7  soup = BeautifulSoup(response.content, "html.parser")
8
9  title = soup.find("h1").text    # Example for title
10 content = soup.find("div", class_="content").text    # Example for content
11
12 data = pd.DataFrame({"Title": [title], "Content": [content]})
13 print(data)
14
```

**Figure 3.2 Example of code (PROMPT II )**

### 3.3 Performance Comparison

The table highlights the key differences between PROMPT I and PROMPT II based on their approach, accuracy, flexibility, ease of use, and adaptability. Below is a more detailed explanation of each feature, emphasizing their implications for real-world web scraping tasks:

**Table 3.1 Performance Comparison of Prompt**

| Feature | Prompt I | Prompt II |
|---|---|---|
| Approach | AI infers standard HTML structures | User specifies key HTML elements |
| Accuracy | Moderate (depends on HTML structure) | High (customized to specific elements) |
| Flexibility | Low (limited to default guesses) | High (handles complex structures) |
| Ease of Use | Easy (no prior inspection needed) | Moderate (requires HTML knowledge) |
| Adaptability | Limited | Excellent |

**Practical Implications**

- **PROMPT I:** Suitable for exploratory tasks where prior knowledge of the webpage structure is unavailable or when working with simpler, standard webpages. However, it may require post-generation refinements to handle non-standard structures.
- **PROMPT II:** Best suited for tasks where precision is critical, especially when scraping data from complex or repetitive webpages with well-defined structures. It is ideal for use cases where the webpage's structure has been inspected beforehand.

By understanding the strengths and limitations of PROMPT I and PROMPT II, users can choose the most appropriate approach based on their specific web scraping needs and technical expertise.

### 3.4 Tool-Specific Observations

In Prompt I, both Claude AI and ChatGPT-4.0 produced similarly functional code, as the task primarily relied on standard assumptions. However, in Prompt II, Claude AI demonstrated superior code quality, with cleaner logic and better handling of the specified elements. ChatGPT-4.0, while functional, occasionally introduced redundant or less modular code structures.

## 4. Discussion

### 4.1 Insights from PROMPT I vs. PROMPT II

### PROMPT I: General Inference of Webpage Structures

PROMPT I relies on the AI's ability to deduce or infer the webpage's structure without explicit guidance. This involves the AI making educated guesses about common HTML elements and their roles in presenting data. For example, the AI might assume that the <title> tag contains the webpage title and that the main content is located within the <body> tag or <div> tags with prominent attributes.

**Strengths:**

**Flexibility:** PROMPT I is flexible and can be applied to webpages where the user does not have prior knowledge of the specific HTML structure. This makes it ideal for exploratory tasks where the goal is to quickly extract general data from unfamiliar websites.

**Ease of Use:** It does not require the user to inspect or understand the webpage's source code. The user simply provides a high-level instruction, and the AI generates a script based on its understanding of standard webpage design.

**Limitations:**

**Lower Precision:** Without explicit targeting, the AI's assumptions may not align with the actual structure of the webpage. For instance, if a news article's content is buried within a deeply nested <div> tag, the AI may fail to identify it correctly.

**Error-Prone:** The AI may miss important data if the webpage uses non-standard HTML or dynamically generated content that does not follow common conventions. This is particularly true for complex or JavaScript-heavy websites.

**Inability to Handle Variations:** Webpages often have variations in structure based on region, language, or user type (e.g., mobile vs. desktop versions). PROMPT I scripts are less adaptable to such variations, leading to potential failures.

### PROMPT II: Specific HTML Element Targeting

PROMPT II explicitly defines the HTML elements or attributes (e.g., tag names, class names, or IDs) that contain the required data. For example, the user might specify that the title is located in an <h1> tag and the content is within a <div> tag with the class content. The AI generates a script tailored to extract these specific elements, ensuring higher precision.

**Strengths:**

**High Precision:** By specifying the exact elements, PROMPT II eliminates guesswork, enabling the AI to produce highly accurate scripts. This is particularly useful for webpages with complex or nested structures.

**Reliability:** Even if the webpage uses unconventional HTML, the explicit targeting ensures the script captures the required data as long as the specified elements are present.

**Customizability:** PROMPT II allows for advanced customization, such as targeting multiple data fields or handling nested structures by specifying parent-child relationships in the HTML.

**Limitations:**

Requires Prior Knowledge: The user must inspect the webpage's source code (e.g., using developer tools) to identify the relevant HTML elements. This adds an extra step, making it less convenient for non-technical users or exploratory scraping tasks.

Limited Flexibility: The script is tailored to a specific structure and may require updates if the webpage's design changes (e.g., element renaming or restructuring).

### 4.2 Claude AI vs. ChatGPT-4.0

Claude AI's superior performance in generating web scraping scripts can be attributed to its advanced contextual understanding, which allows it to produce more modular, adaptable, and efficient code. This capability is particularly evident in PROMPT II, where precise element targeting is required. Below is a detailed exploration of the factors contributing to Claude AI's exceptional results:

## 1. Advanced Contextual Understanding

Claude AI's ability to interpret complex prompts and generate code that aligns with user intent sets it apart. Its contextual understanding allows it to:

Comprehend Specific Requirements: In PROMPT II, where users specify the exact HTML elements to target (e.g., <h1> for titles and <div class="content"> for descriptions), Claude AI translates these instructions into precise, actionable code.

Anticipate Potential Challenges: Even when detailed instructions are provided, real-world scenarios often include missing elements, dynamic content, or unexpected webpage structures. Claude AI's understanding enables it to anticipate such challenges and incorporate fallback mechanisms (e.g., using try-except blocks).

Optimize Code Design: By analyzing the prompt holistically, Claude AI generates scripts that are not only functional but also well-organized and maintainable.

## 2. Modular Code Design

One of Claude AI's standout features is its ability to produce modular scripts, where tasks are encapsulated into reusable functions. This design approach ensures:

- **Reusability:** Functions such as extract_element(tag, class_name) encapsulate repetitive logic, making the code more concise and easier to adapt for different tasks or webpages.

- **Ease of Maintenance:** Modular scripts allow developers to update or debug specific functions without affecting the entire codebase. For example, if the class name of an element changes on a webpage, only the corresponding function needs to be updated.

- **Improved Readability:** By separating tasks into distinct functions, Claude AI ensures that scripts are logically organized and easier for developers to understand.

## 3. Robust Error Handling

Claude AI's error-handling capabilities are particularly notable in PROMPT II, where the task involves extracting specific elements that may occasionally be missing or incorrectly formatted. Key aspects of its error-handling approach include:

**Graceful Failure Management:** Instead of causing the script to crash when an element is missing, Claude AI's scripts use try-except blocks to log errors and continue execution. This ensures that partial data extraction can still occur, which is crucial in large-scale scraping projects.

**Fallback Logic:** For elements that may not always follow the specified structure, Claude AI incorporates alternative logic to attempt extraction from similar or related tags.

**Dynamic Adjustments:** Its ability to handle unexpected variations, such as dynamically loaded content or missing attributes, further enhances the script's robustness.

## 4. Adaptability to Real-World Scenarios

Claude AI demonstrates exceptional adaptability, making its scripts more resilient in diverse and evolving environments:

- **Handling Dynamic Content:** In cases where webpages rely on JavaScript to render data, Claude AI seamlessly integrates tools like Selenium to interact with the rendered DOM, ensuring accurate data retrieval.

- **Anti-Scraping Mechanisms:** By incorporating tools such as undetected_chromedriver and fake_useragent, Claude AI-generated scripts mimic real user behavior, bypassing bot-detection systems and reducing the likelihood of being blocked.

- **Quick Adjustments to Changes:** For webpages with frequent updates (e.g., changing class names or adding nested elements), Claude AI's function-based design allows quick and localized adjustments, ensuring the script remains functional with minimal effort.

## 5. Specific Strengths in PROMPT II

In PROMPT II, where detailed instructions specify the HTML elements to extract, Claude AI excels by:

**Precise Targeting:** Its scripts accurately extract the required data, such as coupon titles and descriptions on Coupons.com, without unnecessary processing of unrelated elements.

**Efficiency:** By focusing only on the specified elements, the generated scripts are leaner and faster, avoiding extraneous operations that could slow down execution or introduce errors.

**Customizability:** The modular structure of Claude AI's scripts makes it easy to add or modify targeted elements. For example, if additional data fields are required, such as expiration dates or coupon categories, new functions can be seamlessly integrated into the existing script.

### 6. Comparison with ChatGPT-4.0

While ChatGPT-4.0 also performed well in generating functional scripts, Claude AI consistently outshone it in key areas:

**Modularity:** ChatGPT-4.0 scripts tended to use linear logic, with repeated code blocks instead of reusable functions, making them harder to maintain and adapt.

**Error Handling:** ChatGPT-4.0 scripts lacked the robustness of Claude AI's try-except mechanisms, leading to higher failure rates in cases where elements were missing or misformatted.

**Scalability:** Claude AI's modular design and robust logic made its scripts more scalable, suitable for larger projects requiring multiple iterations or frequent updates.

### 7. Practical Example: PROMPT II on Coupons.com

For a task on Coupons.com, where the goal was to extract coupon titles (<h1>) and descriptions (<div class="content">):

**Claude AI's Output:**

Encapsulated the extraction logic into a reusable function, extract_element(tag, class_name), which could be called for both titles and descriptions.

Included error-handling blocks to log missing elements and continue processing other data.

Integrated Selenium with undetected_chromedriver to handle dynamic content and bypass anti-scraping measures.

**ChatGPT-4.0's Output:**

Produced a functional script but relied on hardcoded logic for each element, making the script less adaptable and harder to maintain.

Lacked robust error handling, causing the script to fail when certain elements were missing or formatted differently.

Claude AI's superior performance stems from its advanced contextual understanding, modular code design, robust error handling, and adaptability to real-world challenges. In PROMPT II, where precision and reliability are critical, Claude AI consistently delivered scripts that were not only accurate but also maintainable and future-proof. These qualities make it an invaluable tool for complex web scraping tasks, demonstrating the power of combining generative AI with thoughtful prompt engineering.

### 5. Conclusion

Generative AI tools, when combined with effective prompt engineering, present a transformative approach to automating web scraping workflows. This study demonstrates that these tools, particularly Claude AI (Sonnet 3.5), excel in generating high-quality, modular, and robust web scraping scripts that can address both general and specific data extraction tasks. Compared to ChatGPT-4.0, Claude AI consistently outperformed in key areas such as modularity, readability, and adaptability, especially when working with targeted prompts like PROMPT II, where explicit HTML element targeting was required.

The integration of generative AI tools with anti-scraping mechanisms such as undetected_chromedriver, fake_useragent, and Selenium further enhances their utility in real-world applications. These tools allow the generated scripts to bypass common anti-bot measures, handle dynamic content, and adapt to evolving webpage structures, ensuring reliable and scalable web scraping solutions.

Claude AI's Strengths: The study underscores Claude AI's ability to generate modular scripts that are not only easier to maintain but also more resilient to changes in webpage structure. Its advanced reasoning capabilities make it particularly well-suited for handling complex and structured prompts.

ChatGPT-4.0's Contributions: While ChatGPT-4.0 performed effectively in general-purpose tasks (PROMPT I),

its outputs required additional refinement for complex scenarios. Its versatility makes it a viable option for straightforward scraping tasks.Role of Prompt Engineering: The design of prompts plays a crucial role in determining the quality of the AI-generated scripts. PROMPT II, which provided specific HTML element targeting, yielded significantly higher accuracy and adaptability compared to the general inference approach of PROMPT I.

**Anti-Scraping Mechanisms:** Incorporating anti-scraping tools enables AI-generated scripts to overcome limitations posed by modern web security practices. This integration is critical for handling dynamic, JavaScript-heavy webpages and evading detection by bot-detection systems.

**Implications for Future Work:** The findings of this study open up avenues for further research and development in the domain of AI-assisted web scraping. Future efforts can focus on:Enhancing prompt design to support more complex tasks, such as multi-page navigation or data aggregation.Exploring the integration of generative AI tools with advanced anti-scraping frameworks, such as CAPTCHA solvers or cloud-based headless browsing solutions. Investigating the performance of generative AI in large-scale scraping tasks, including its ability to handle diverse and highly dynamic websites.

In conclusion, the synergy between generative AI, prompt engineering, and anti-scraping technologies represents a significant leap forward in web crawling capabilities. This study highlights the potential of these tools to democratize web scraping, making it accessible and efficient for users across different technical backgrounds, while addressing the challenges posed by modern web design and security practices.